\documentclass[preprint,showpacs,showkeys,superscriptaddress,aps,pra,12pt]{revtex4-2}

\usepackage{graphicx}
\usepackage{epstopdf}
\usepackage{bm}
\usepackage{amssymb}
\usepackage{amsmath}
\usepackage{bbold}
\usepackage{ulem}
\usepackage{color, soul}
\usepackage{xcolor}
\usepackage{afterpage}
\usepackage{footmisc}
\usepackage{natbib}
\usepackage{float}
\usepackage{multirow}

\begin{document}
	
	
	\title{Spin order dependent skyrmion stabilization in MnFeCoGe hexagonal magnets}
	\author{Dola Chakrabartty}%
	\affiliation{School of Physical Sciences, National Institute of Science Education and Research, HBNI, Jatni-752050, India}
    \author{Mihir Sahoo}%
	\affiliation{RPTU Kaiserslautern-Landau, Kaiserslautern, Germany}
	\author{Amit Kumar}%
	\affiliation{Solid State Physics Division, Bhabha Atomic Research Centre, Trombay, Mumbai 400085, India}
	\author{Sk Jamaluddin}%
	\affiliation{School of Physical Sciences, National Institute of Science Education and Research, HBNI, Jatni-752050, India}
	\author{Bimalesh Giri}%
	\affiliation{School of Physical Sciences, National Institute of Science Education and Research, HBNI, Jatni-752050, India}
	\author{Hitesh Chhabra}%
	\affiliation{School of Physical Sciences, National Institute of Science Education and Research, HBNI, Jatni-752050, India}
	\author{Kalpataru Pradhan}%
	\email{Kalpataru.Pradhan@saha.ac.in}
	\affiliation{Theory Division, Saha Institute of Nuclear Physics, A CI of Homi Bhabha National Institute, Kolkata-700064, India}
	\author{Ajaya K. Nayak}
	\email{ajaya@niser.ac.in}
	\affiliation{School of Physical Sciences, National Institute of Science Education and Research, HBNI, Jatni-752050, India}
	\affiliation{Center for Interdisciplinary Sciences (CIS), National Institute of Science Education and Research, HBNI, Jatni, 752050 Bhubaneswar, India.}
	
	\date{\today}

\begin{abstract}

\textbf{Topological magnetic skyrmions in centrosymmetric systems  exhibit a higher degrees of freedom in their helicity, hence possess a great potential in the advanced spintronics including skyrmion based quantum computation. However, the centrosymmetric magnets also display non-topological trivial bubbles along with the topological skyrmions.  Hence it is utmost priority to investigate the impact of different magnetic ground states and their underlying interactions on  the stabilization of magnetic skyrmions in cetrosymmetric magnets.   Here, we present a combined theoretical and experimental study on the role of non-collinear magnetic ground state   on the skyrmion stabilization in a series of exchange frustrated non-collinear ferromagnetic system MnFe$_{1-x}$Co$_x$Ge. With the help of neutron diffraction (ND) and Lorentz transmission electron microscopy (LTEM) studies, we show that hexagonal skyrmions lattice emerges as a stable field driven state only when the underlying magnetic ground state is  collinear with easy-axis anisotropy. In contrast, non-topological type-II bubbles are found to be stable state in the case of  non-collinear magnetic ordering with partial in-plane anisotropy.   Furthermore, we also find that the skyrmions transform to the non-topological bubbles when the system undergoes a spin reorientation transition from the easy-axis to easy-cone ferromagnetic phase. Our results categorically establish  the significant role of in-plane magnetic moment/anisotropy  that hinders the  stability of skyrmion both in the case of collinear and non-collinear magnets. Thus, the present study offers a wide range of opportunities to manipulate the stability of dipolar skyrmions by changing the intrinsic characteristics of the materials.}	
\end{abstract}

\maketitle

\section*{\textbf{INTRODUCTION}}

Non-collinear magnetic textures with topological character, such as skyrmions are one of the major interests in spintronics community due to their enormous scientific and technological possibilities in the future generation data storage devices with higher density and low power consumption \cite{ultralowcurrentdensity, fege ultralow current, emergent electrodynamics, Tokura_review}. At the early stages of skyrmion discovery, focus was mostly concentrated on bulk and thin film magnets with broken inversion symmetry for hosting Dzyaloshinskii-Moriya interaction, the major force for the stabilization of skyrmions \cite{MnSi, FeCoSi, Cu2OSeo3, CoZnMn, GaVs, blowingskyrmion, Ir/Fe/Co/Pt, Neelskx, Tokura_review}. However, in recent years, a wide variety of materials preserving the  inversion symmetry are found to facilitate skyrmion like topological spin textures with different helicity and vorticity \cite{LSFMO, Fe3Sn2, LSMO0P175, MnNiGa, Mn4Ga2Sn, NdMn2Ge2, NdCo5}. These centrosymmetric systems provide additional degrees of freedom to the  helicity and vorticity of the skyrmions, which offers greater flexibility for their implementation as "0" and "1" data bits in storage device and quantum computing  \cite{helicitybit, helicity2, helicity1, helicity3, helicity4, helicity5}. In  most of these magnets competing  uniaxial anisotropy and dipolar energy  play a major role in the stabilization of skyrmions \cite{LSFMO, Fe3Sn2, LSMO0P175, MnNiGa, Mn4Ga2Sn, NdMn2Ge2, NdCo5}. Recently, a few studies have demonstrated that the modification in uniaxial anisotropy and/or application of external in-plane magnetic field can greatly influence the stability of dipolar skyrmions \cite{type-IIbubble, inplane_fe3sn2, Mn4Ga2Sn, Fe3Sn2_tiltedanisotropy}.  Hence, it is extremely important to investigate the impact of different energy parameters on the stability of skyrmions in centrosymmetric magnets. In this direction, the present manuscript focuses on the role of competing exchange interactions on the stability of magnetic skyrmions. 

As uniaxial magnetic anisotropy (UMA) is one of the key requirements for skyrmion formation in centrosymmetric magnets, we look into systems having potential to exhibit both UMA and competing exchange interactions.  In this direction, the  Ni$_2$In crystal structure based hexagonal magnet MnFeGe is reported to show  non-collinear magnetic structure driven by competing ferromagnetic (FM) and antiferromagnetic (AFM) exchange interactions \cite{MnCoGe,MnFeCoGe2}. It has also been theoretically demonstrated that the magnetic ordering in the MnFeGe system can be modified by altering the inter-atomic distance between the Mn atoms sitting at different layers \cite{MnFeCoGe2}. Furthermore, it has also been shown that a complete replacement of  Fe by Co  leads to a collinear ferromagnetic ground state in the hexagonal magnet MnCoGe \cite{MnCoGe2}.  Hence, it is expected that the substitution of Co can systematically change the landscape of different magnetic interaction in the system.

\begin{figure*}
	\begin{center}
		\includegraphics[width= 15cm]{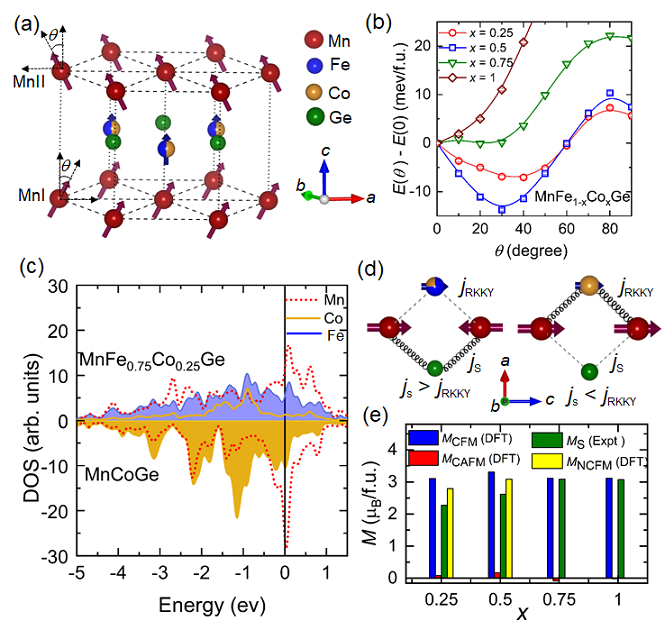}
		\caption{(a) Schematic representation of noncollinear magnetic structure for  MnFe$_{1-x}$Co$_x$Ge . $\theta$ represents the canting angle of Mn moments with respect to the $c$-axis. (b) The change in energy for MnFe$_{1-x}$Co$_x$Ge [$x$ = 0.25. 0.5, 0.75, 1] samples with the canting angle $\theta$. (c) Projected density of state (DOS) for d-states of Mn, Fe and Co for non-magnetic MnFe$_0.75$Co$_0.25$Ge and MnCoGe systems. (d) The schematic representation of possible indirect exchange interactions between Mn-Mn localized moments with changing Co concentration. (e) A comparison of theoretically predicted net magnetic moments for different potential magnetic configurations with experimentally measured saturation magnetizations ($M_S$). For comparison the $M_S$ value of nearby compositions i.e $x$ = 0.3, 0.5, and 0.8 is used. The $M_S$ value for $x$ = 1 is taken from reference \cite{MnCoGe}}   
		\label{fig1}
	\end{center}
\end{figure*}

In order to gain a deep insight into the possible magnetic ground states in the Co doped samples, we utilize first principles density functional theory (DFT) calculation, as implemented in Vienna ab-initio simulation package (VASP) \cite{VASP1, VASP2},  for carrying  out  magnetic structure optimizations and energy calculations by substituting Co in place of Fe  in MnFe$_{1-x}$Co$_x$Ge.  The Perdew-Burke-Ernzerhof (PBE) exchange-correlation functional, a version of generalized gradient approximation (GGA), is considered in the calculations \cite{VASP3}. To investigate the effect of doping concentration on magnetic ground state, we have taken 1$\times$1$\times$2 supercells for all the calculations. In the unit cell, Mn, Fe/Co and Ge atoms occupy the position at (0, 0, 0), (2/3, 1/3, 1/4) and (1/3, 2/3, 1/4) sites, respectively. The k-point grid of 5$\times$5$\times$3 is used to sample the first Brillouin zone for self-consistent calculations. The threshold value for energy convergence between two consecutive electronic relaxation steps is set to 1$\times$10$^{-5}$~eV and the structures are optimized until the force on each atom becomes less than 0.001~eV/Å. The unit cell of MnFeGe structure belongs to the Cm space group with optimized lattice constants of a = b = 4.096 Å, and c= 5.083 Å, which is consistent with previous theoretical results \cite{MnFeCoGe2}. First we compare the total energy differences between the FM and
AFM states $\Delta$$E$ [= $E$(FM) -$E$(AFM)] for MnFeGe and MnCoGe (where all Fe atoms from MnFeGe are replaced by Co atoms) compounds. For MnFeGe, $\Delta$$E$ = 0.067 eV, where Mn-Mn couple antiferromagnetically. But, FM configuration is more stable ($\Delta$$E$ = -0.048 eV) for MnCoGe case.

Interestingly, for partial Co doped MnFeGe system, namely MnFe$_{0.75}$Co$_{0.25}$Ge (MnFe$_{0.5}$Co$_{0.5}$Ge) $\Delta$$E$ decreases to 0.026 eV (0.015 eV), which is quite small. On the other hand, the ground state is FM ($\Delta$E = -0.011 eV) for Co rich system MnFe$_{0.25}$Co$_{0.75}$Ge like that of MnCoGe system. Therefore, our collinear magnetic calculations point to a strong competition between the underlying magnetic states in MnFe$_{1-x}$Co$_x$Ge compounds. So, we extend our calculations to unveil the possible noncollinear magnetic structures. The schematic representation of the expected noncollinear magnetic ordering with out-of-plane FM component and in-plane AFM component is depicted in Fig.~\ref{fig1}(a).  The change in calculated energy as a function of  canting angle ($\theta$) of the Mn moment with respect to the $c$-axis for the samples MnFe$_{1-x}$Co$_x$Ge [$x$ = 0.25, 0.5, 0.75, 1] is shown in Fig.~\ref{fig1}(b). In the case of Co doped sample with $x$ = 0.25, a local energy minima is observed at $\theta$ = 40$^{\circ}$, which nearly matches with the earlier reported $\theta$ $\approx$ 45$^{\circ}$ for the parent compound MnFeGe \cite{MnFeCoGe2}.  Importantly, a further increase in the Co concentration to $x$ = 0.5 leads to the noncollinear ferromagnetic state with $\theta$ = 30$^\circ$ as minimum energy state. The Co rich samples with $x$ = 0.75 and $x$ = 1 show the energy minimum for collinear FM state with $\theta$ = 0$^{\circ}$. Hence, our theoretical study clearly points toward the emergence of non-collinear magnetic state for the sample $x$ = 0.5 and a change in magnetic ordering from the noncollinear AFM to collinear FM ground state with high Co doping ($x$ $\geq$ 0.75) in the MnFe$_{1-x}$Co$_x$Ge system.

To get more insight into the magnetic phase transition with increasing the Co doping, we calculate the density of states (DOS) and projected DOS (PDOS) for the systems. The PDOS for d-states of Mn, Fe and Co for non-magnetic MnFe$_{0.75}$Co$_{0.25}$Ge and MnCoGe systems are shown in Fig.\ref{fig1}(c). A significant difference in DOS near the Fermi level for both the materials in terms of Fe-d and/or Co-d states is clearly observed.  These variations in the Fermi-level states can be related to the indirect exchange model of magnetic ordering, i.e. competition between carrier mediated (RKKY-like) exchange and superexchange \cite{DOS1}.  The parameters of these exchange interactions varies inversely with the energy required to push an electron from d-states to the Fermi level \cite{DOS1, DOS2, DOS3}. Therefore, the amount and distribution of DOS adjacent to Fermi level plays a crucial role in determining these exchange parameters. In the case of sample with high Fe concentration, the number of available Fe-d states is more above the Fermi level than just below it. In contrast, there are more Co-d states present just below the Fermi level in case of MnCoGe compound. In this case, the difference between the DOS of Fe and Co can be compared to the rigid-band model, in which the addition of an extra electron causes the Fermi level to shift upward \cite{DOS1}. Using a perturbative approach, the coupling constants ($j_{RKKY}$ and $j_S$ ) can be expressed in q→0 limit as follows: \begin{equation}
	j_{RKKY}(0) = V^4 D(\epsilon_F)/E^2_h
	\end{equation} \begin{equation}
j_{S}(0) = V^4 \sum_{nk}^{\epsilon_{nk} > \epsilon_F} (\epsilon_{nk} - \epsilon_F -E_h)^{-3}
	\end{equation} Here, $V$ and $D(\epsilon_F)$ denote electron mixing parameter and DOS at the Fermi level, respectively. $\epsilon_{nk}$ is the energy at k-point of the nth band whereas $E_h$ is the energy in electron
	transfer from d-states to the Fermi level. Although both MnFe$_{0.75}$Co$_{0.25}$Ge and MnCoGe display a large Co-d states at the Fermi level, a sharp fall in DOS (mainly in terms of Fe-d
		states) just above the Fermi level is found when all the Fe atoms are replaced by Co. Thus, the number of unoccupied states (N) near the Fermi level is very small in case of MnCoGe than Fe rich case. Thus, the superexchange coupling constant $j_S$ stated in the above equation is bounded from above by $j_S$(0)$\le$ $V^4N$ /$E_h$$^3$ \cite{DOS1, DOS2, DOS3}. Therefore Fe rich system ($|$${j_S}$$|$$>$$|$$j_{RKKY}$$|$ due to the large N ) prefers AFM ordering. In the case of MnCoGe, N is smaller compared to that of MnFe$_{0.75}$Co$_{0.25}$Ge and hence RKKY coupling constant $j_{RKKY}$ should be greater than the superexchange constant $j_S$. In addition, it is also
	noted that $D(E_F)$ for MnCoGe is larger than that of MnFe$_{0.75}$Co$_{0.25}$Ge, hence in overall, FM ordering in MnCoGe. Figure~\ref{fig1}(d) schematically represents the dominant RKKY exchange mediated through the Co conduction electron as a possible origin of ferromagnetism rather than antiferromagnetic ordering, in the Co rich samples. In fact, the spin-polarized site-projected
		DOS for Co atom for MnCoGe in FM configuration (Figure not shown here) shows
		a sharp peak in spin-down DOS arising from Co-d states just below the Fermi level. This
		state helps in mediating the indirect interactions between Mn atoms favoring a FM ground
		state for MnCoGe \cite{ kalpstsru_sir_1, kalpstsru_sir_2}. The calculations also indicate that the observed non-collinearity for the intermediate Co compounds, such as MnFe$_{0.5}$Co$_{0.5}$Ge, is mainly driven by the competition between different types of indirect exchange interactions between the localized Mn moments. 

This suggests that the MnFe$_{1-x}$Co$_x$Ge system is a potential candidate for hosting tunable magnetic ground states depending on the nature of spin-ordering. Therefore,  in this report, we focus on the evolution of magnetic skyrmions and bubbles with change in energy landscape   in the series of centrosymmetric magnets MnFe$_{1-x}$Co$_x$Ge. With help of Lorentz transmission electron microscopy (LTEM) study, we demonstrate how the presence of inplane magnetic anisotropy hinders the formation of skyrmion lattice in the case of non-collinear magnets ($x$ $\le$ 0.6) as well as easy-cone FM state,  whereas the same can be easily stabilized in the case of easy axis collinear FM state in the sample $x$ = 0.8.


\section*{RESULTS AND DISCUSSION}

\begin{figure*}
	\begin{center}
		\includegraphics[width= 16cm]{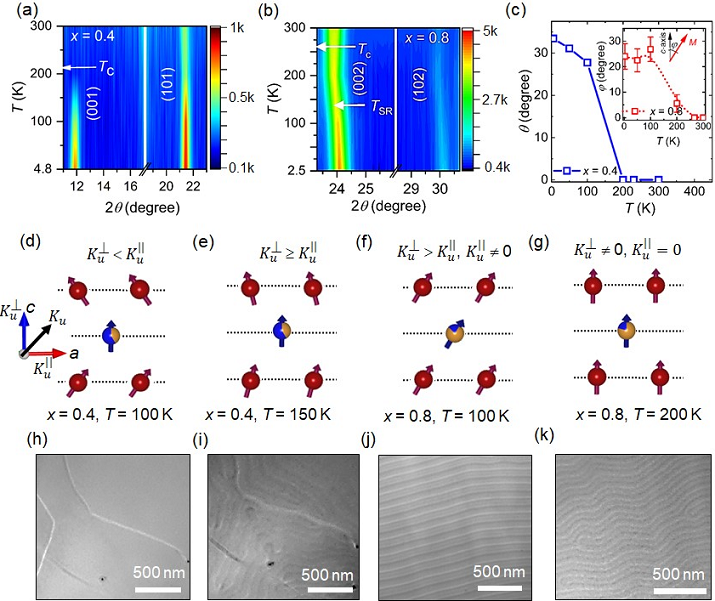}
		\caption{Contour plot of powder neutron diffraction (PND) profile intensity for the samples (a) $x$ = 0.4, and (b) $x$ = 0.8. The color bars represent the profile intensity of PND data. (c) The temperature dependent canting angle ($\theta$) of Mn moments with respect to the $c$-axis for the sample $x$ = 0.4. The inset shows temperature dependent tilting of magnetic easy axis ($\phi$) with respect to $c$-axis for the sample $x$ = 0.8.  (d)-(g) Schematic representations of magnetic ground states obtained from the PND measurements for $x$ = 0.4 and 0.8.  Corresponding zero field magnetic domains observed using over-focused Lorentz transmission electron microscopy (LTEM) for  (h) $x$ = 0.4 and $T$ = 100~K, (i) $x$ = 0.4 and $T$ = 150~K, (j) $x$ = 0.8 and $T$ = 100~K,  and (k) $x$ = 0.8 and $T$ = 200~K.}  
		\label{fig2}
	\end{center}
\end{figure*}
\begin{figure*}
	\begin{center}
		\includegraphics[width= 17cm]{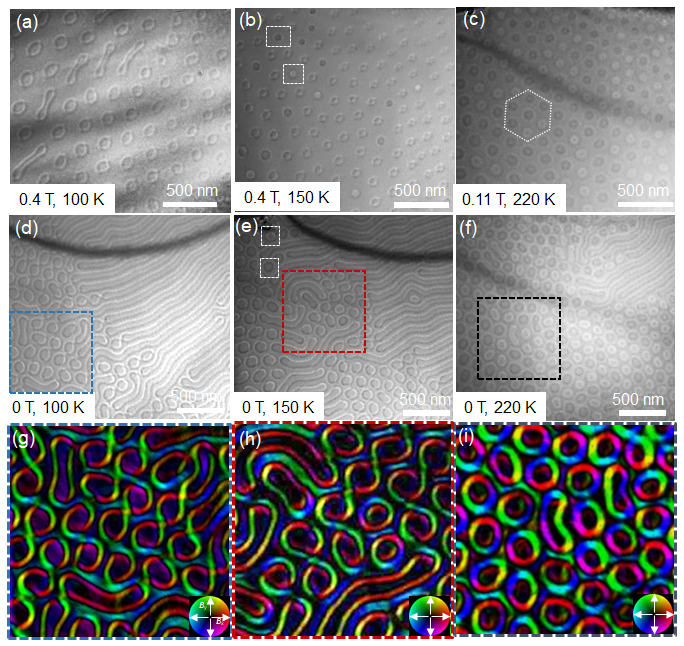}
		\caption{Temperature evolution of magnetic domains for the samples MnFe$_{0.2}$Co$_{0.8}$Ge.  The over-focused LTEM images at (a) $ T =  $100~K, (b)  $ T =  $150~K, and (c)  $ T =  $ 220~K. In all the cases, the magnetic field is applied along the $c$-axis. The hexagonal skyrmion lattice is marked with hexagon and the isolated skyrmions are marked with doted boxes. (d)-(f) The temperature evolution of the remnant magnetic states obtained after increasing the magnetic field along zone axis to the skyrmion/bubble state and then reducing the field to zero. (g)-(i) Real space spin textures constructed using transport of intensity equation (TIE) analysis of the marked regions in (d)-(f). The color wheels represents direction of the in-plane magnetization components. }  
		\label{fig3}
	\end{center}
\end{figure*}
Polycrystalline samples of  MnFe$_{1-x}$Co$_x$Ge (x = 0.2 to 0.8) are prepared using arc melting technique.  All the samples formed in the layered hexagonal crystal structure with space group P6$_3$/mmc. 
Our magnetic measurements show an increase in the magnetic ordering temperatures ($T_C$)  from 170~K to 260~K with increasing Co doping from $x$ = 0.2 to $x$ = 0.8 . We also find a monotonic increase in the saturation magnetic moment ($M_S$) from 2~$\mu_B$/f.u. to 3.1~$\mu_B$/f.u with increasing Co concentration from $x$ = 0.2 to $x$ = 0.8.  Furthermore, we have compared the theoretically calculated magnetic moments of different magnetic configurations with the experimentally obtained saturation magnetization ($M_S$), as shown in Fig.~\ref{fig1}(e). The $M_S$ matches well with the DFT calculated total FM moment for the samples with $x$ = 0.75 and 1, whereas  for $x$ = 0.25 and $x$ = 0.5, the experimental $M_S$  better matches with the total calculated moment for the non-collinear ferromagnetic (NCFM) configuration rather than collinear FM configuration.

To further experimentally verify the theoretically proposed change in the magnetic ground states in the present system,  powder neutron diffraction (PND) measurement is carried out on two of our samples with $x$ = 0.4 and $x$ = 0.8 (Co rich) as shown in Fig.~\ref{fig2}(a) and Fig.~\ref{fig2}(b), respectively. In the case of $ x=0.4 $, a substantial rise in the intensity for (001) and (101) reflections below $T_C$ can be seen clearly in the contour plot in Fig.~\ref{fig2}(a). In addition, the appearance of (001) magnetic reflection only below the $T_C$ indicates the presence of a finite basal plane AFM component \cite{MnCoGe, afmnc2, afmnc3}. 
To get an idea about the degree of noncollinearity in case of $ x=0.4 $, we have plotted the temperature dependent canting angle ($\theta$) of the Mn moments  obtained  from the PND refinement [Fig.~\ref{fig2}(c)]. At $ T= $ 4.8~K, the canting angle is nearly 33.4$^{\circ}$~$\pm$~0.8$^{\circ}$, which matches well with the theoretically predicted $\theta$ for $x$ = 0.5 sample [see Fig.~\ref{fig1}(b)]. The $\theta$ gradually decreases with increasing temperature and becomes nearly 0$^{\circ}$ around the $T_C$. For $x$ = 0.8 sample,  the PND data do not show any additional magnetic reflections other than the ones on  top of the nuclear reflections. The findings suggest the presence of a collinear magnetic ordering for $x$ = 0.8 sample \cite{MnCoGe}, as suggested by our theoretical calculations. Furthermore, a close analysis of the PND data reveals an increase in the strength of the (002) and (102) reflections with decreasing temperature below 150 K, as shown in the contour plot Fig.~\ref{fig2}(b). To understand this peculiar behavior,  we have simulated the PND data for this sample with the magnetic easy-axis tilted away from the $c$-axis by an angle $\phi$ \cite{srtpnd1, srtpnd2, srtpnd3}.  This results in a monotonic enhancement of the (002) and (102) reflections with  increasing $\phi$. Hence, we have carried out the Rietveld refinement of the PND data for this sample with the easy-cone model.   The temperature variation of refined tilting angle ($\phi$)  for the $x$ = 0.8 sample is shown in the inset of Fig.~\ref{fig2}(c). We find a tilting angle of about 25$^{\circ}$~$\pm$~5$^{\circ}$ below 100~K. Therefore, the sample $x$ = 0.8 exhibit an easy-cone FM state below 150~K and an easy-axis ordering at higher temperatures. 

\begin{figure*}
	\begin{center}
		\includegraphics[width= 16cm]{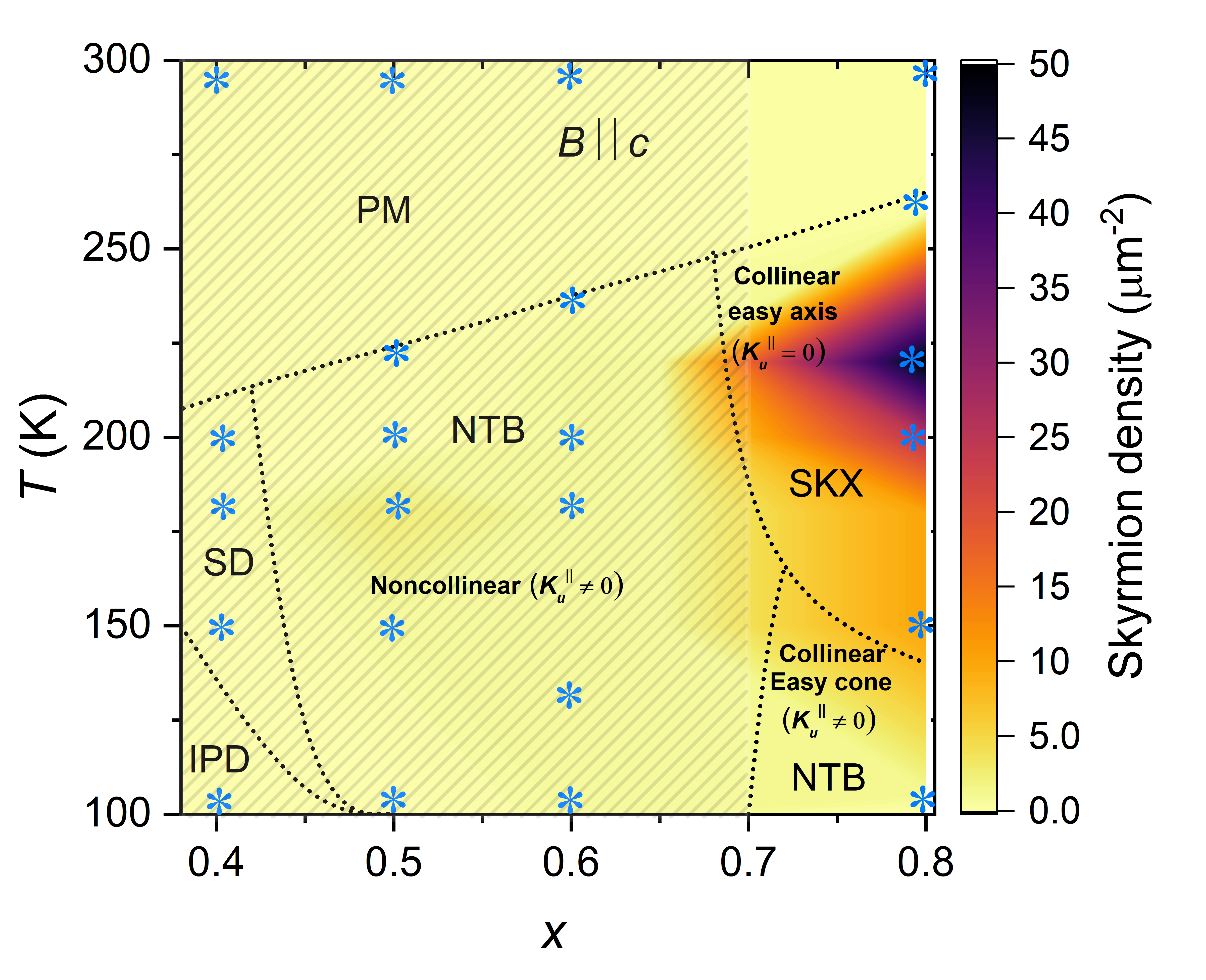}
		\caption{A temperature ($T$) vs. Co concentration ($x$) phase diagram for the samples MnFe$_{1-x}$Co$_x$Ge. The Color bar represent the skyrmion density per micrometer square area. The LTEM data at a magnetic field $H$ is used, where the maximum number of skyrmion are observed for that particular $T$ and $x$. SKX, SD, IPD, NTB, PM represent skyrmion lattice, stripe domain, in-plane domain, type-II bubble, and paramagnetic phase, respectively. * symbol represents the LTEM data points.}  
		\label{fig4}
	\end{center}
\end{figure*}

The schematic representation of the magnetic ground states for the samples $x$ = 0.4 and 0.8 obtained from the PND data  along with the corresponding real space LTEM images recorded at zero external field are shown in Fig.~\ref{fig2}(d)-(k). As depicted in Fig.~\ref{fig2}(d), the presence of non-collinear magnetic state with a large  in-plane anisotropy component ($K_{u}^\parallel$) gives rise to the observation of  in-plane magnetic domains  as spontaneous magnetic state  for the sample $x$ = 0.4 at 100K [see Fig.~\ref{fig2}(h)]. Furthermore, due to a decrease in the $K_{u}^\parallel$ in comparison  to the out-of-plane magnetic anisotropy  ($K_{u}^\perp$) [see Fig.~\ref{fig2}(e)], the magnitude of the in-plane domain walls starts diminishing and an impression of the out-of-plane stripe domains starts appearing as a spontaneous state [see Fig.~\ref{fig2}(i)]. Interestingly, the sample with $x$ = 0.8, which displays an easy cone magnetic state at 100~K [see Fig.~\ref{fig2}(f)],  exhibits a stripe domain as spontaneous magnetic  state as shown in Fig~\ref{fig2}(j). When the $K_{u}^\parallel$ component completely vanishes at $T =$ 200~K for $ x= $ 0.8 [see Fig.~\ref{fig2}(g)], we also find the presence of  little bit disordered stripe domains [Fig.~\ref{fig2}(k)] compared to that observed at $T =$ 100~K.
This type of stripe domain alignment in case on in-plane anisotropy has previously been found experimentally by introducing a finite $K_{u}^\parallel$ component to the system \cite{Fe3Sn2_tiltedanisotropy}.  Additionally, our Object-Oriented Micromagnetic Framework (OOMMF)-based micromagnetic simulations also demonstrates that the stripe domain can be organized in the direction of the in-plane anisotropy component with the tilting of the magnetic easy-axis. 

To further study the nature of magnetic state we have carried out a detailed field and temperature dependent LTEM study in the present system. In the case of $ x= $ 0.4, the in-plane domain state at 100~K [Fig~\ref{fig2}(h)] and the mixed state at 150~K [Fig~\ref{fig2}(i)] transform to field polarized phase with the application of magnetic field. However, the scenario changes completely in the case of $x$ = 0.8, where we find different domain states by varying the magnetic field and temperature.   To eliminate the effect of in-plane magnetic field, all the LTEM experiments are carried out with the applied magnetic field  along the $c$-axis. As shown in Fig.~\ref{fig3}(a), the Co rich sample $x$ = 0.8 exhibits a hexagonal lattice of type-II bubbles at 100~K and a magnetic field of 0.4~T.  By increasing the temperature to 150~K leads to a mixed state of type-II bubble and skyrmions at a magnetic field of 0.4~T [see Fig.~\ref{fig3}(b)]. The presence of higher number of type-II bubbles than that of skyrmions  suggest that former are the energetically stable state at this temperature. Surprisingly, we find the stabilization of only skyrmions with both clockwise (CW) and counter-clockwise (CCW) helicity by increasing the temperature to 220~K [see Fig.~\ref{fig3}(c)]. It is important to mention here that our PND clearly show the existence of a collinear FM ground state with easy-axis anisotropy at 220~K, whereas an easy-cone FM arrangement with tilted easy-axis with respect to the $c$-axis is found at 100~K. Hence, it is very much evident that the stabilization of tupe-II bubbles greatly depend on the nature of magnetic ordering in the system.

It has been also reported that the presence of small in-plane applied magnetic field can break the symmetry of the spin alignment in a centrosymmetric skyrmion, thereby giving rise to the observation of type-II bubbles \cite{Mn4Ga2Sn, type-IIbubble}.  To confirm that the present observations are free from the effect of in-plane magnetic field, we have recorded the zero field remnant magnetic state after initially applying the field exactly along the zone axis and then decrease the field to zero. The remnant magnetic states at different temperatures for the sample $x$ = 0.8 are shown in the Fig.~\ref{fig3}(d)-(f). At 100~K a mixed phase of type-II bubbles and stripe domains is observed [see Fig.~\ref{fig3}(d)], suggesting a very small energy difference between these magnetic states. As expected at 150~K, a few skyrmions along with the magnetic state of 100~K  are found [see Fig.~\ref{fig3}(e)]. 
On the other hand, the observation of mixed phase of skyrmions and stripe domains without the existence of any type-II bubbles at 220~K suggests that the skyrmions are having lower energy in the system compared to the type-II bubbles [see Fig.~\ref{fig3}(f)].  To confirm the exact nature of spin arrangements in the observed magnetic structures,  we have  constructed the spin textures using transport of intensity equation (TIE) analysis of the marked regions. The TIE analysis of the LTEM images clearly shown the presence of Bloch-type skyrmions at 220~K, whereas the existence of Bloch-point type feature in the domain state at 100~K and 150~K indicates the formation of type-II bubbles.

Although the stripe domains are observed as spontaneous magnetic state for the sample $x$ = 0.4 at 150~K, no skyrmion state is observed with the field evolution of the magnetic domains. 
Here, the point should be noted that the sample with $x$ = 0.4 has a non-collinear FM ground state with sufficiently strong in-plane AFM component (i.e, $K_{u}^\parallel$ $\ne$ 0 ) in the basal plane. For the sample $x$ = 0.5, 0.6 a hexagonal type-II bubble lattice along with a very few number of skyrmions are observed. 
 Figure~\ref{fig4} shows Co concentration ($x$) vs. temperature ($T$) phase diagram for the samples MnFe$_{1-x}$Co$_x$Ge. Here, the LTEM images at the magnetic field that hosts the maximum number of skyrmion at a specific ($x$, $T$) are used to determine the skyrmion density for that point. The strong skyrmion density is observed for the easy axis collinear ferromagnet with zero $K_{u}^\parallel$, whereas the samples with lower Co concentration upto $x$ = 0.6 shows mostly type-II bubbles as stable magnetic state. Furthermore, the easy-cone FM phase with finite $K_{u}^\parallel$ exhibits type-II bubbles as a stable state rather than skyrmion. All the experimental observations demonstrates that the presence of in-plane anisotropy component in a system hinders the skyrmions stability. All the experimental observations are also theoretically validated using object-oriented micromagnetic framework. the simulated data shows that the skyrmions can be transformed to the type-II bubble by introducing a sufficient amount of in-plane magnetic anisotropy component along with the out-of-plane anisotropy.            

The skyrmion-like textures  in the uniaxial centrosymmetric systems are of great technological interest due to their  different topological numbers as well as helicity degrees of freedom.  In most cases, the competing UMA and dipolar interaction are considered as the fundamental mechanisms for skyrmion stabilization in the centrosymmetric system \cite{LSFMO, Fe3Sn2, LSMO0P175, MnNiGa,NdCo5, NdMn2Ge2, Mn4Ga2Sn, Fe3GeTe2}. Furthermore, some of the centrosymmetric systems show skyrmions of size 1-2~nm  due to frustrated magnetic interaction including four spin exchange interaction \cite{Gd2PdSi3, GdRu3Al12, GdRu2Si2}. However, these extremely small skyrmions are always found at very low temperatures  ($<$~10~K). In this direction, the addition of frustrated magnetic exchange  to the dipolar skyrmion systems might serve as an important step forward  to realize small skyrmions at room temperature.  The dipolar stabilized skyrmions in most of the centrosymmetric systems  are always considered in the collinear ferromagnetic  backgrounds. Although, few of the earlier literatures describe the tunablity of the dipolar skyrmions in terms of external stimuli, such as magnetic field \cite{Mn4Ga2Sn, Fe3GeTe2, type-IIbubble, inplane_fe3sn2} and current \cite{Fe3Sn2_current control, Fe3Sn2_APL, Fe3sn2_CURRENT2}, their stability while modifying the internal energy parameters  is not thoroughly investigated.  Moreover, the effect of magnetic ground states and the underlying interactions on the dipolar stabilized skyrmions is still remain elusive. In the present report, a comprehensive investigation on the stability of dipolar skyrmions depending on the strength of exchange frustration and the corresponding magnetic ground states is demonstrated, where the frustration in the magnetic exchange interactions between Mn moments can be tuned depending on the Fe and Co atomic ratio \cite{MnFeCoGe2, MnCoGe, MnCoGe2}. 

Our theoretical calculations and experimental findings indicate the presence of a noncollinear canted magnetic ground state for the samples MnFe$_{1-x}$Co$_x$Ge with $x$$<$0.75 and a collinear ferromagnetic state when $x$$\gneq$0.75. The LTEM observation of in-plane domain walls for the non-collinear ferromagnet with $x$ = 0.4 at $T$ = 100~K supports the presence of a higher in-plane magnetic anisotropy component ($K_u^\parallel$), correlating to a significant in-plane AFM component. As a result, it is expected that the  out-of-plane ($K_u^\perp$) and in-plane ($K_u^\parallel$) magnetic anisotropy components realign based on the change in the out-of-plane or in-plane magnetic moment contributions. The DFT calculations, on the other hand, indicate a decrease in the canting angle ($\theta$) in the noncollinear magnets with increasing Co concentration ($x$). Hence a decrease in the ratio of $K_u^\parallel$/$K_u^\perp$ can lead to the emergence of out-of-plane stripe domains as a zero field LTEM state. Most importantly, in noncollinear background ( $K_u^\parallel$ $\ne$ 0) the non-topological type-II bubbles are mainly stabilized as field driven stable state rather than the topological skyrmions. The hexagonal skyrmion lattice is only observed in collinear ferromagnetic background when the magnetic field applied along the easy-axis direction. Earlier reports show a transformation between the topological skyrmions and non-topological type-II bubbles in the collinear ferromagnetic background with application of non-zero in-plane magnetic field \cite{Mn4Ga2Sn, type-IIbubble}. The present study elucidate  that the skyrmions in collinear ferromagnetic background can also be transformed to type-II bubble with an applied magnetic field along the zone axis, when a nonzero in-plane magnetic moment or $K_u^\parallel$ introduced in the system. All our experimental findings point that the $K_u^\parallel$ inhibits skyrmion stability in both collinear and non-collinear magnetic backgrounds. Hence, the present study sheds light on the consequences of different energy factors on the stability and tunability of dipolar skyrmions.

\section*{\textbf{CONCLUSION}}

To summarize, we have thoroughly investigated how the magnetic ground states and the corresponding interactions affect the stability of dipolar skyrmions in a variety of non-collinear hexagonal ferromagnets MnFe$_{1-x}$Co$_x$Ge. We show that the degrees of non-collinearity and the exchange frustration strength possess a significant correlation with the stability of dipolar skyrmions. The skyrmion lattice can only be stabilized in the easy-axis collinear ferromagnet with applied magnetic field along the zone axis, whereas non-topological type-II bubbles emerges as more favorable state in the non-collinear magnetic background. The role of in-plane magnetic anisotropy, $K_u^\parallel$, in the skyrmion stabilization is demonstrated in the case of easy-cone  ferromagnetic phase where  type-II bubble are found.  Furthermore, our research provides the prospect of broad control over dipolar skyrmions by modifying the internal energy characteristics of the materials, and it might be regarded as a step ahead in the realization of dipolar skyrmion-based spintronic devices.



\section*{Acknowledgments}
 AKN acknowledges the support from Department of Atomic Energy (DAE), the Department of Science and Technology (DST)-Ramanujan research grant (No. SB/S2/RJN-081/2016), SERB research grant (ECR/2017/000854) and Nanomission research grant
[SR/NM/NS-1036/2017(G)] of the Government of India. A.K. Nayak thank Amitabh Das, Bhabha Atomic Research Centre, Mumbai for recording the powder neutron diffraction
data.

\end{document}